\documentclass{ws-procs975x65}

\begin{document}

\title{Parity Doublet Model applied to Neutron Stars}

\author{V. Dexheimer$^*$, S. Schramm, H. Stoecker}

\address{FIAS, Johann Wolfgang Goethe University,\\
Frankfurt am Main, Germany\\
$^*$dexheimer@th.physik.uni-frankfurt.de}

\begin{abstract}

The Parity doublet model containing the SU(2) multiplets including the baryons identified as the chiral partners of the nucleons is applied for neutron star matter. The chiral restoration is analyzed and the maximum mass of the star is calculated.  
\end{abstract}

\keywords{Neutron Star, Chiral Symmetry, Parity Model}

\bodymatter

\section{The model}

The doublet parity model assumes, besides the nucleons, the presence of particles with opposite chirality named chiral partners. At high density environment as the one in neutron stars interior, there is the possibility of creating such heavy particles. The main difference between this model and the usual chiral model (Ref.~\refcite{chi}) is that in this one there is a term of bare mass in the lagrangian density. This mass, called $M_O$ appears with fields that are a mixture of the fields of the particles and their chiral partners in such a way that it does not break chirality (Ref.\refcite{par}).

It is assumed that the star is in chemical equilibrium and the baryons interact through the mesons $\sigma$, $\omega$ and $\rho$ (included in order to reproduce the high asymmetry between neutrons and protons). Electrons are included to insure charge neutrality. The lagrangian density of the system contains besides the kinetic and the interaction terms an explicit symmetry breaking term in order to reproduce the masses of the pseudo-scalar mesons. The coupling constants of the baryons are adjusted to reproduce the baryonic vacuum masses. In the high-density limit the nucleon and its chiral partner have degenerate masses ($M_0=790 MeV$) as the sigma field goes to zero and chiral symmetry is restored:

\begin{eqnarray}
M^*_\pm=\sqrt{\left[\frac{(M_{N_+}+M_{N_-})^2}{4}-M_0^2\right]\frac{\sigma^2}{\sigma_0^2}+M_0^2}\pm\frac{M_{N_+}-M_{N_-}}{2}\frac{\sigma}{\sigma_0}.
\end{eqnarray}

A good candidate for the nucleon chiral parter is the N'(1535), but since the identification of chiral partners is still on its first steps we study the case with N'(1200) and N'(1500) for comparison. This variation has drastic consequences on the results. Four different cases first studdied in Ref.~\refcite{par} are applied to neutron stars.

\section{Results} 

When the density increases, the protons, the neutron and the proton chiral partners appear, respectively. Since the matter is not symmetric, there is no need for the chiral partners to appear at the same rate. The transition to the chiral phase is abrupt for the cases with the fourth order autointeraction term of the vector mesons $g_4=0$ (P2 and P4) because of the sudden appearance of the chiral partners (Fig.~\ref{massp}(a)).

\def\figsubcap#1{\par\noindent\centering\footnotesize(#1)}

\begin{figure}[b]
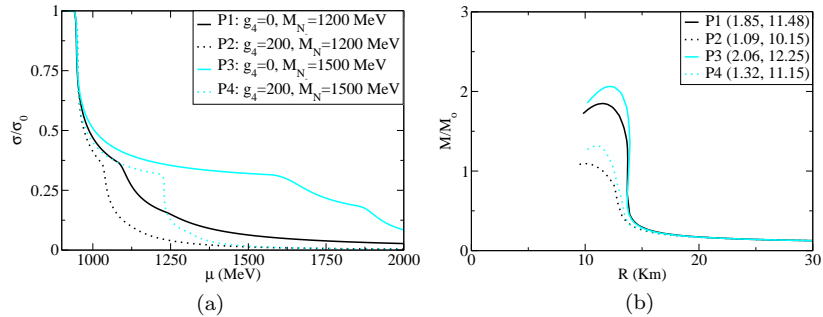
%
\begin{center}
  \parbox{2.1in}{\epsfig{figure=sigma.eps,width=2.15in}\figsubcap{a}}
  \hspace*{4pt}
  \parbox{2.1in}{\epsfig{figure=massp.eps,width=2in}\figsubcap{b}}
  \caption{(a)Scalar condensate versus chemical potential. (b)Star mass versus radius.}%
  \label{massp}
\end{center}
\end{figure}

The calculus of the mass of the star shows that bigger maximum values are reached for $g_4=0$. That means that when $g_4$ increases, the value of the vectorial meson $\omega$, related to it, decreases its value and consequently its repulsive effect in such a way that the star can hold smaller quantity of mass against collapse (Fig.~\ref{massp}(b)).

\section{Conclusion}

With increasing density .i.e. toward the center of the star,  chiral partners begin to appear, reaching a point where they exist at the same  rate as they corresponding particles. The decrease in the scalar condensates signals the restoration of the chiral phase. Depending on the parameters, the phase transition turns out to be a continuous cross-over or of first order for P4.

The maximum mass of the star is higher when the coupling constant $g_4$ is set to zero (P1 and P3) so in order to reproduce the most massive star observed, that has $M=2.1^{+0.4}_{-0.5}M_{\odot}$ (Ref.~\refcite{21}), the best option would be the P1 description, because although the P3 predicts a higher maximum mass for the star, its phase transition occurs at a chemical potential bigger than 1700 MeV, too hight compared to predictions.

\end{document}